\documentclass[Letter]{emulateapj}
\usepackage{apjfonts}

\usepackage{epsfig}

\newcommand{\arcm}{\hbox{$^\prime$}}

\newcommand{\degree}{\hbox{$^\circ$}}
\newcommand{\rosat}{\emph{ROSAT}}
\newcommand{\chandra}{\emph{Chandra}}
\newcommand{\xmm}{\emph{XMM-Newton}}
\newcommand{\xmms}{\emph{XMM}}
\newcommand{\asca}{\emph{ASCA}}
\newcommand{\einstein}{\emph{Einstein}}
\newcommand{\arcs}{\mbox{\arcm\arcm}}
\newcommand{\Lx}{\ensuremath{L_{\mathrm{X}}}}

\newcommand{\Zsol}{\ensuremath{Z_{\odot}}}

\newcommand{\Msol}{\ensuremath{M_{\odot}}}

\newcommand{\s}{\ensuremath{\mbox{~s}}}
\newcommand{\ps}{\ensuremath{\s^{-1}}}

\newcommand{\erg}{\ensuremath{\mbox{~erg}}}
\newcommand{\ergps}{\ensuremath{\erg \ps}}

\newcommand{\Msmbh}{\ensuremath{M_{\mathrm{SMBH}}}}

\begin{document}

\title{AGN feedback and gas mixing in the core of NGC~4636}

\author{E. O'Sullivan\altaffilmark{1}, J. M. Vrtilek\altaffilmark{1} and
  J. C. Kempner\altaffilmark{2}}
\altaffiltext{1}{Harvard-Smithsonian Center for Astrophysics, 60 Garden
  Street, Cambridge, MA 02138, email: \textit{eosullivan, jvrtilek@head.cfa.harvard.edu}}
\altaffiltext{2}{Dept. of Physics and Astronomy, Bowdoin College, 8800
  College Station, Brunswick, ME 04011, email: \textit{jkempner@bowdoin.edu}}

\shorttitle{AGN feedback in NGC~4636}
\shortauthors{O'Sullivan et al}

\begin{abstract}
  \chandra\ observations of NGC~4636 show disturbances in the galaxy X-ray
  halo, including arm-like high surface brightness features (tentatively
  identified as AGN driven shocks) and a possible cavity on the west side
  of the galaxy core. We present \chandra\ and \xmm\ spectral maps of
  NGC~4636 which confirm the presence of the cavity and show it to be
  bounded by the arm features. The maps also reveal a $\sim$15 kpc wide
  plume of low temperature, high abundance gas extending 25-30 kpc to the
  southwest of the galaxy. The cavity appears to be embedded in this plume,
  and we interpret the structure as being entrained gas drawn out of the
  galaxy core during previous episodes of AGN activity. The end of the
  plume is marked by a well defined edge, with significant falls in surface
  brightness, temperature and abundance, indicating a boundary between
  galaxy and group/cluster gas. This may be evidence that as well as
  preventing gas cooling through direct heating, AGN outbursts can produce
  significant gas mixing, disturbing the temperature structure of the halo
  and transporting metals out from the galaxy into the surrounding
  intra-group medium.
\end{abstract}

\keywords{galaxies: individual: NGC 4636 --- galaxies: intergalactic medium --- galaxies: halos --- X-rays: galaxies}

\section{Introduction}

NGC~4636, the dominant galaxy of a group on the outskirts of the Virgo
cluster, is one of the most well-known nearby ellipticals. It is also one
of the most X-ray luminous, and has been studied in detail by every major
X-ray imaging observatory to date. Spectral imaging with \einstein\ showed
the galaxy to be surrounded by an extensive halo of hot gas with a
temperature of 0.78-1.21 keV \citep[90\% error
bounds,][]{Formanjonestucker85}. Further studies using \rosat\ and \asca\ 
measured the metal abundance of the gas, found gradients in both the
temperature and abundance with radius, and showed the halo to be very
extended
\citep{Awakietal94,Trinchierietal94,Matsushitaetal97,FinoguenovJones00,Buote00,OPC03}.
More recently, NGC~4636 has been observed by both \chandra\ and \xmm. The
initial \chandra\ ACIS-S observation showed unusual structures in the core
of the galaxy, most notably ``spiral arm'' features and a possible ring of
high surface brightness emission, thought to be the product of shocks
driven by a previous AGN outburst \citep{Jonesetal02}. Detailed spectral
analysis of these data suggests the presence of a cavity in the X-ray halo
on the west side of the core \citep{Ohtoetal03}. Again, AGN activity is the
likely cause; plasma from the radio jets of the AGN can displace the gas of
the X-ray halo, leaving an apparent void and altering the projected X-ray
properties at that point. Analysis of \textit{XMM} RGS spectra shows that
there is little or no emission from gas at temperatures below 0.5 keV, and
therefore no cooling flow in the system \citep{Xuetal02}. NGC~4636 hosts a
weak radio source (1.4$\times$10$^{38}$ \ergps) with small scale jets which
account for $\sim$40\% of the radio emission at 1.4 Ghz
\citep{BirkinshawDavies85,StangerWarwick86}. The H$\alpha$ luminosity is
measured to be 2.3$\times$10$^{38}$ \ergps\ \citep{Hoetal97}, and an
upper limit of 2.7$\times$10$^{38}$ \ergps\ is estimated for the X-ray
emission of the AGN \citep{Loewensteinetal01}.

Using the high quality data available from both \chandra\ and \xmms, we
present spectral maps of the inner galaxy halo of NGC~4636, which reveal
interesting temperature and abundance structures. We consider the origin of
these structures and their impact on our understanding of galaxy and group
halos.

\begin{figure*}
\centerline{\includegraphics[angle=-90,width=18cm]{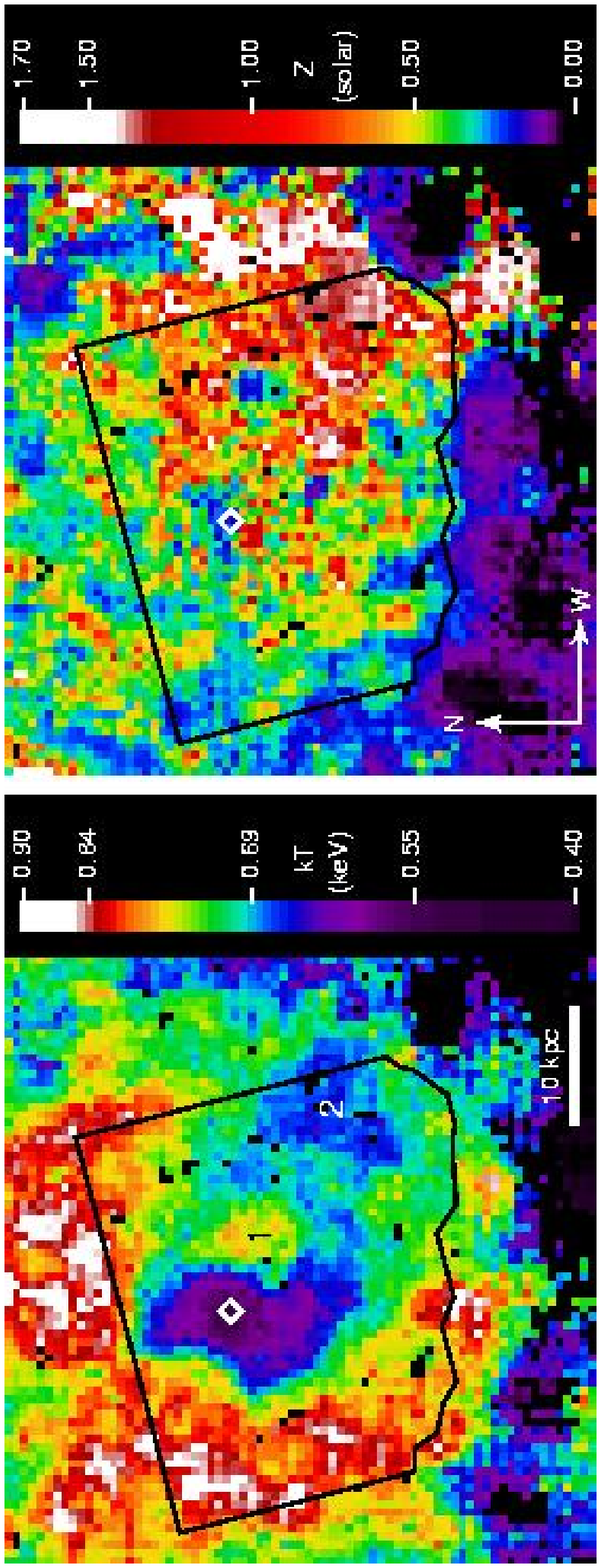}}
\centerline{\includegraphics[angle=-90,width=18cm]{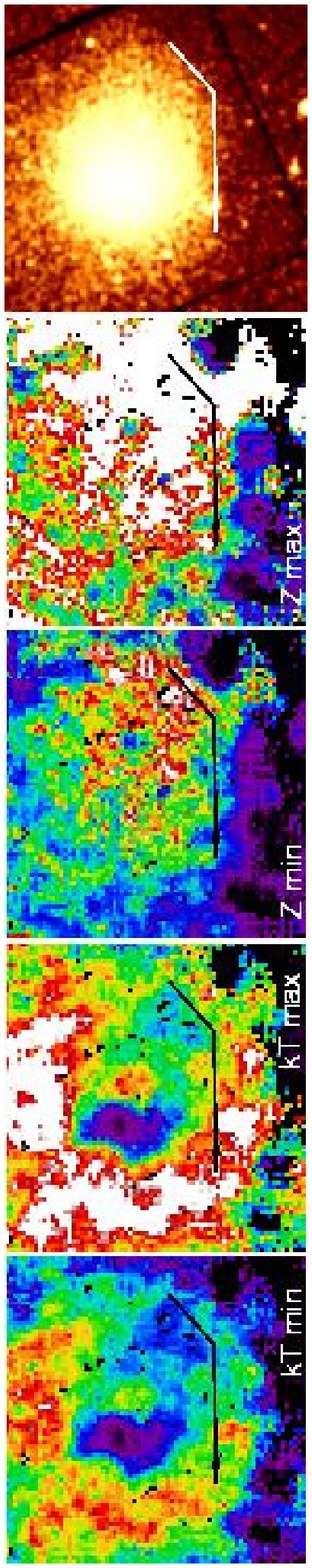}}
\caption{\label{fig:xmm_maps} \textit{Upper panel}: \textit{XMM} spectral
  maps of the southwest part of the halo of NGC~4636, showing
  (\textit{left}) Temperature in keV and (\textit{right}) Abundance in
  solar units. The maps use 75$\times$73 8.8\arcs$^2$ pixels, whose
  spectral extaction regions are squares of side 9-67.5\arcs$^2$.  The
  diamonds on each map mark the center of the galaxy, regions marked 1 and
  2 are, respectively, the cavity and plume referred to in the text. Black
  lines mark the region of overlap with the \chandra\ spectral maps (see
  Figure~\ref{fig:cha_maps}).  \textit{Lower panel}: Error maps showing the
  90\% upper and lower limits on temperature and abundance in each pixel,
  and a gaussian smoothed MOS 1 surface brightness image of the same field.
  Lines mark the southern surface brightness edge referred to in the text.
  10 kpc is equivalent to $\sim$130\arcs.}
\vspace{-7mm}
\end{figure*}

\section{Observations and Data Analysis}
NGC~4636 was observed with the \chandra\ ACIS-S (chip S3) for 53 ks on 2000
January 26-27 (ObsID 323) and with \xmms\ EPIC for 64 ks on 2001 January
5-6 (ObsID 0111190701). For the \chandra\ observation, the instrument
operated in faint mode, and reduction and analysis of the data were
performed using \textsc{ciao} v.3.1 and CALDB v.2.28. Reprocessing, removal
of background flares, and creation of responses and background were carried
out as described in \citet{OSullivanPonman04b}. Point sources were
identified using the \textsc{wavdetect} tool. Apart from those in the
galaxy core, which were considered false, all were removed.

The \xmms\ EPIC instruments were operated in full frame mode with the
medium optical filter. The raw data were processed using \textsc{sas}
v.6.0, and reprocessing, filtering for bad pixels and columns and times of
high background, and point source removal were performed as described in
\citet{OSullivanetal03}, including use of the `(FLAG == 0)' filter.
Background events lists were generated from the \citet{ReadPonman03}
blank-sky background data, cleaned to match the data.  The \textsc{sas
  evigweight} task was applied to both source and background events lists,
``correcting'' the events to account for vignetting.  This allows the use
of single, on-axis ARF response files for each instrument, which were
generated using \textsc{arfgen}. Pregenerated RMF files were used, selected
by position.  The exposure times after cleaning were $\sim$58.4 ks for the
MOS cameras, $\sim$50.9 ks for the PN, and $\sim$38.9 ks for ACIS-S.

\begin{figure*}
\centerline{\includegraphics[angle=-90,width=18cm]{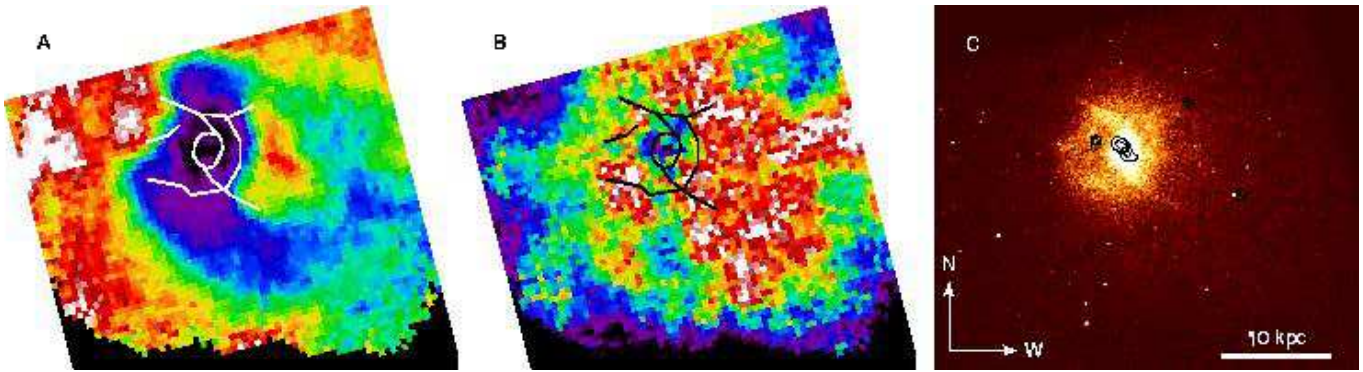}}
\caption{\label{fig:cha_maps} \chandra\ temperature (\textit{A}) and
  abundance (\textit{B}) maps of the core of NGC~4636, using the same color
  scheme as in Fig.~\ref{fig:xmm_maps}. The maps use 64$\times$64
  6.4\arcs$^2$ pixels, whose spectral extaction regions are squares of side
  8.4-63\arcs$^2$. For comparison, a \chandra\ 0.5-2.0 keV surface
  brightness (\textit{C}) image is also shown. All images are on the same
  scale and centered on the same coordinates. VLA 21cm radio emission
  contours are overlaid on panel \textit{C}, and outlines of the surface
  brightness features are overlaid on panels \textit{A} and \textit{B} to
  aid comparison. 10 kpc is equivalent to $\sim$130\arcs.}
\vspace{-7mm}
\end{figure*}

In order to look for correlations between the structures visible in the
X-ray surface brightness images and temperature and abundance changes in
the halo, we created spectral maps from the \chandra\ and \xmms\ datasets
using the method described in \citet{OSullivanetal05}. The spectra for each
map pixel were required to have $>$800 counts, and were fitted with an
absorbed MEKAL model with hydrogen column fixed at the galactic value
(1.8$\times$10$^{20}$ cm$^{-2}$). Energies below 0.35 (0.5) keV and above
6.0 (8.0) keV were ignored in the \chandra\ (\xmms) fits. Pixels whose 90\%
errors on kT were $>$10\% (15\%) were excluded from the \chandra\ (\xmms)
maps. Figure~\ref{fig:xmm_maps} shows temperature and metal abundance maps
generated from the \xmms\ data, with associated 90\% error maps. For
comparison, Figure~\ref{fig:cha_maps} shows \chandra\ temperature and
abundance maps, plotted using the same color scale as used in
Figure~\ref{fig:xmm_maps}, as well as an X-ray surface brightness image.
In each figure, all panels are plotted on the same scale and coordinates,
so as to be directly comparable.

Several important features are visible in the temperature maps. Perhaps the
most obvious is a region of low temperatures (blue-purple) in the galaxy
core and extending in a curve to the south and southwest. To the west of
the galaxy core is an area of higher temperatures apparently surrounded by
cool gas (region 1 in Figure~\ref{fig:xmm_maps}). To the west of this is a
large area of moderately cool temperatures (region 2) apparently connected
to the core on both sides of region 1. To the north, east, and southeast of
the core the temperatures are higher, and this high temperature gas appears
to form a boundary around the cool galaxy core and regions 1 and 2, though
at the western edge of region 2 it is appears to be relatively narrow, with
cooler gas beyond. Three main features can be drawn from the abundance
maps; first that there is a boundary between high abundance gas in the
inner halo, and low abundance gas at the southern and western edges of the
maps, with a similar boundary between high and low temperatures; second
that the highest abundances are seen to the southwest of the galaxy core,
near region 2 of the temperature map; third, that the abundances in the
galaxy core appear to be only $\sim$0.5 solar.

Comparing these features to the X-ray surface brightness images we find
that the high temperature region 1 seems to be bounded by the western part
of the ``spiral arm'' features. In particular, the southwest ``spiral arm''
marks the boundary between region 1 and the low temperature curve extending
from the galaxy core (see Figure~\ref{fig:cha_maps}). On larger scales the
surface brightness appears roughly circular (see
Figure~\ref{fig:xmm_maps}), but a sharp drop in surface brightness along
the south and southwest sides of the inner halo is visible, coincident with
the end of region 2 and the southern edge of the area of high abundance.
This is probably the boundary between the galaxy halo and a surrounding
intra-group or -cluster medium (ICM).  Regions 1 and 2 lie roughly along
the minor axis of the stellar body of the galaxy. The AGN radio jets do not
extend as far as region 1, and are initially aligned $\sim$43\degree away
from the minor axis of the galaxy \citep{BirkinshawDavies85}, but at a
small distance from the center they appear to bend, with the southern jet
turning to point toward region 1.

There are issues which could affect the absolute accuracy of our maps. Many
map pixels will not be independent, due to overlapping spectral extraction
regions and the large \xmms\ PSF. We have also not corrected for
differences in the soft component of the background between the source and
background datasets. The fact that both \chandra\ and \xmms\ maps show the
same features and similar temperatures and abundances at given positions is
a strong argument that the PSF issue and background differences are not
causing significant problems. It is also important to point out that we are
interested in relative differences in temperature and abundance across the
maps, rather than the absolute value in each pixel. However, as a further
test, we carried out more traditional spectral fits for selected regions of
the \xmms\ data. We extracted spectra from regions centered on the features
in the spectral maps and fitted them with a variety of models. With the
exception of the galaxy core, all fits produced results in agreement with
the maps. The galaxy core required a multi-temperature plasma and power-law
model, and while the resulting temperatures agreed with the maps, abundance
was a factor $\sim$2 higher. The cause is probably the ``Fe-bias'' effect,
the single temperature map fit underestimating the abundance of
multi-temperature emission \citep{Buotefabian98,Buote00b}.  With this
exception, we believe these spectral fits show the maps to be an accurate
representation of the true temperature and abundance structure.

\section{Discussion} 
\label{sec:discuss}

The high temperatures observed at region 1 correspond closely with the
results reported by \citet{Ohtoetal03}, and confirm their suggestion of a
cavity in the gas. The apparent curve of the southern radio jet toward
region 1 adds further support, and we note that the cavity lies along the
minor axis of the galaxy, where the density gradient might be assumed to be
steepest. The temperature map suggests that the cavity is non-spherical,
and instead has a concave surface toward the galaxy center. This is
reminiscent of the bell-like shapes seen in some numerical simulations of
buoyantly rising bubbles of radio plasma in galaxy clusters
\citep[e.g.,][]{Churazovetal01}. Given the high temperatures around the
north, east, and southeast of the core, it seems likely that the core is
surrounded by a spherical shell of hot gas through which a plume of cool
gas, ending in region 2, has penetrated. Under this assumption, the high
temperature of the cavity is a projection effect, indicating that there is
less cool gas along the line of sight than is the case for the surrounding
regions.

Taking a simple model of the cavity as a spherical bubble of radio plasma
expanding into and rising through the X-ray plasma of the galaxy halo, we
can estimate the time taken to create the cavity. Assuming the cavity
expands without causing shocks in the halo, we can limit the expansion
velocity to the sound speed of the halo gas, leading to an expansion time
of $\sim$23 Myr. From \citet{Churazovetal00}, the velocity at which the
bubble rises owing to its own buoyancy is $v_b=C\sqrt{rGM/R^2}$, where
\textit{R} is the distance of the bubble from the galaxy core, \textit{r}
is the radius of the cavity, and C is a numerical constant for which a
value of 0.5 is accepted for incompressible fluids. This yields a buoyant
rise time of $\sim$47.5 Myr, but this should be considered a lower limit as
this assumes a large difference in gas density between the galaxy core and
the current location of the bubble.  These two timescales together suggest
that the bubble may have formed close to its current location, rather than
at the core of the galaxy.  We can also estimate the power required to
produce the cavity, assuming it expands at the sound speed and the pressure
equilibrium between the radio and X-ray plasmas is maintained. From
\citet{Churazovetal00}, the power required is $Power=4\pi r^3KP/3t$ where
$K=(\gamma-1)/\gamma$, $\gamma$ is the adiabatic index of the relativistic
gas in the bubble (i.e. $\gamma$=4/3), \textit{t} is the expansion time of
the cavity and \textit{P} is the pressure of the halo around the cavity.
From this we estimate the power required to be $\sim$4.8$\times$10$^{42}$
\ergps. This can be compared to the current luminosity of the core of
NGC~4636, \Lx$\simeq$3$\times10^{41}$\ergps\ within 20~kpc.

These estimates are considerably higher than the observed X-ray and radio
emission of the AGN, but 2-3 orders of magnitude below estimates of the
power required to create the larger cavities seen in some galaxy clusters
\citep{Bohringeretal02}. The radio luminosity to jet power conversion
factors of \citet{Bicknelletal97} suggest that the total current jet power
is $\lesssim3\times10^{41}$\ergps\ (assuming a magnetic field strength
$>$10$^{-4}$G and the age of the jet to be the bubble expansion timescale
), and given the lack of a direct connection between the jets and the
cavity, it seems likely that the AGN is currently in a quiescent state, and
that the cavity is a remnant of a previous outburst.  A similar situation
is observed in Abell~4059, which has large cavities in the X-ray halo, but
a central radio galaxy whose has declined since their formation
\citep{Heinzetal02}. There is also the possibility that the AGN outburst
was more powerful than our calculations suggest, and caused supersonic
expansion of the bubble; the ``spiral arm'' features in the core were
\citet{Jonesetal02} identified by \citet{Jonesetal02} as shocks, and a
rapidly expanding cavity might shock surrounding gas.  However, the
resolution of our spectral maps and the size of the spectral extraction
regions means that we cannot comment on these features, except to point out
their association with the cavity. In any case, the cavity size and
power required are small by the standards of galaxy clusters.

One possible reason for this could be that AGN power is related to the mass
of the central supermassive black hole (SMBH), so that differences in SMBH
mass (\Msmbh) cause differences in outburst strength and cavity size.
\citet{MerrittFerrarese01} estimate \Msmbh=7.9$\times$10$^7$ \Msol\ for
NGC~4636, based on the stellar velocity dispersion, while
\citet{vanderMarel99} calculates \Msmbh=3.6$\times$10$^8$ \Msol\ based on
the central optical brightness profile. Estimates for M87 are higher,
\Msmbh=3.0$\times$10$^9$ \Msol\ \citep{Tremaineetal02} being typical. This
suggests that NGC~4636 has a SMBH $\sim$5-40 times less massive than M87,
and if outburst power is directly proportional to \Msmbh, this might
explain the difference in cavity sizes. It is unclear whether there is a
general difference in \Msmbh\ between group and cluster dominant
ellipticals, as only a few measurements are available. However, we note
that in the \citet{MerrittFerrarese01} sample, all galaxies with
\Msmbh$>$10$^9$ \Msol\ are cluster dominant ellipticals, apart from
NGC~4649, which is at the center of a subclump of the Virgo cluster.
Another point is that galaxy groups (and individual ellipticals) have short
cooling times compared to clusters. This implies that the AGN duty cycle in
NGC~4636 should be short; after an outburst a cooling flow will be rapidly
reestablished and may soon fuel a new outburst. These factors suggest a
model in which NGC~4636 produces small outbursts relatively often, while
cluster dominant ellipticals produce a smaller number of more powerful
events.

Region 2 may be relevant to this argument. The temperature maps suggest
that there is a plume of gas extending southwest from the galaxy core, with
region 2 forming the tip of the plume and the cavity of region 1 embedded
within it. The direction of this plume matches that of the galaxy minor
axis. The high abundances to the southwest of the galaxy core suggest that
much of the gas in this area has been enriched, and the most likely
explanation of this is that the gas was enriched inside the galaxy, then
transported outward. In such a model, the plume and cavity can be
interpreted as a snapshot of an ongoing series of small AGN outbursts,
producing cavities which entrain metal rich, cool gas from the galaxy core.
Entrainment of gas by buoyant bubbles is predicted from simulations
\citep{Churazovetal00}, and high resolution X-ray observations of the
radio/X-ray arms of M87 show regions of multi-temperature gas which seem
likely to be entrained gas \citep{Molendi02}. The width of these structures
in M87 is $\sim$8-10 kpc, whereas the NGC~4636 plume
is $\sim$15 kpc across.

If this scenario is correct, the cool plume and high abundances require the
AGN of NGC~4636 to have been active prior to the episode which produced the
cavity visible in the temperature map, and to have significantly affected
the structure of the galaxy halo. The removal of cool, enriched gas from
the galaxy core would clearly affect the formation of any cooling flow, and
could have important implications for enrichment in galaxy groups, as it
provides a mechanism for the diffusion of metals from the central
elliptical into the surrounding intra-group medium (IGM). The sharp drop in
X-ray surface brightness, temperature and abundance at the south and west
edges of the maps all suggest that this marks the boundary between gas
associated with the galaxy, enriched by stellar mass loss and supernovae,
and gas associated with a larger potential, which has a much lower
metallicity.  This surrounding gas could be the halo of the group of which
NGC 4636 is the dominant member, or the Virgo ICM.  Confirmation of ICM
with kT$\simeq$0.5 keV and 0.1-0.2\Zsol\ abundance at 2.6 Mpc from the
cluster core would be an interesting result in itself, but may be
unrealistic.  Simulations of entrainment by AGN jets in clusters suggest
that the mixing effect is not strong enough to affect observed abundance
gradients \citep{Bruggen02}. However, these simulations do suggest that gas
can be moved distances of 15-20 kpc, comparable to the motion suggested by
our maps. If there is a boundary in NGC~4636 between gas associated with
the galaxy and the IGM at a distance of 25-30 kpc, then motions on the
scale seen in the simulations would be sufficient to mix the two,
particularly if multiple AGN outbursts are considered.

\section{Conclusions}
Using X-ray data from \xmm\ and \chandra, we have mapped the temperature
and metallicity structure of the inner halo of NGC~4636. The maps confirm
the presence of a cavity to the west of the galaxy, and show a plume of
cool, metal--rich gas extending beyond the cavity to the southwest. Both
cavity and plume appear to be the product of past AGN activity, the AGN
being quiescent at present. The most likely scenario involves AGN outbursts
producing bubble which then entrain enriched gas as they rise buoyantly
from the galaxy core. If this is the case, then AGN activity may play an
important role in producing both the temperature and abundance structures
we observe in galaxy groups.

\acknowledgments The authors would like to thank L. David, T. Ponman and C.
Jones for useful discussions of this Letter, and an anonymous referee for
several helpful comments. Support for this work was provided by NASA grant
AR4-5012X.

\end{document}